# Structural aspects of deformation defects in bulk metallic glasses (I)*

Yang Tong

Department of Materials Science and Engineering, University of Tennessee, Knoxville, TN 37996, USA

**Mechanical behaviors of bulk metallic glasses (BMGs) including heterogeneous and homogeneous deformation are interpreted by phenomenological shear transformation zones (STZs) model. Currently, information about STZs, i.e. size and density, is only extracted by fitting model equation to the data obtained from macroscopic mechanical tests. This is inadequate since structural features of STZs theory cannot be assessed. Here, we develop anisotropic pair distribution function (PDF) method for directly characterizing mechanical response of deformation defects. Our results reveal the physical picture of deformation defects in BMGs and also provide direct experimental observation of a link between mechanical deformation and intrinsic properties of deformation defects in BMGs.**

Structure-mechanical deformation relationship in metals and alloys is well understood because of unambiguously identified plastic deformation carrier such as dislocation. Unlike dislocation in metals and alloys, lack of well-defined defects in BMGs makes the study of structural origin of plastic deformation particularly challenging. Currently, the broadly accepted plastic deformation theory of BMGs is phenomenological STZs theory proposed by Argon[1,2]. STZs theory provides reasonable atomic mechanisms of heterogeneous deformation[1] at room temperature (RT) and homogeneous deformation[1,2] at high temperature (HT), but experimental verification of these atomic mechanisms by directly characterizing STZs is still missing. The lack of studies on direct characterization of STZs in real space by TEM and even HRTEM is due to inherent features of

STZs: a) shear transformation in STZs is activated by external stresses and has limited time-scale[3]; b) atoms in STZs have no clear structural contrast with atoms in surrounding matrix[4].

Existing methods of STZs characterization are indirect and rely on macroscopic or microscopic mechanical tests, such as HT compression[3,5,6] and RT nanoindentation[7-9]. To estimate STZ size and density, data obtained from those mechanical tests are fitted by formula[1,8,10] derived from a STZ model. Using nanoindentation technique, Pan et al. found that BMGs with higher Poisson's ratio, $\nu$, have larger STZ size[7] and better ductility. This finding, however, is not general since there are brittle Pd-based BMGs[11] with $\nu$ higher than some ductile Zr-based BMGs[12]. Moreover, high temperature mechanical tests estimate that STZs contains 20~30 atoms[3,5,6]. Other RT nanoindentation investigations estimate that STZs include 20~30 atoms[8,9] while values around 200~700 atoms[7] are also claimed. These STZ size values, however, without doubt mislead us to conclude that STZ size is constant or even decreases with temperature, which is shown wrong later in this paper. Additionally, the size of STZ estimated by these indirect characterization methods is questionable since several parameters are tuned in the fitting formula. Also, STZ density is ignored in their formulas.

Here, we demonstrate that anisotropic component of the PDF[13,14] is an accurate method of characterizing both defects' size and density because it describes local atomic re-arrangements in response to stress in terms of inter-atomic position in real space. More importantly, this method firstly provides a direct evidence of defects' structure without relying on STZ model fitting. In the following section, the theoretical basis of anisotropic PDF is discussed first, and then utilizing this framework, we study stress, composition, pressure, and temperature dependence of defects to reveal the structural origins of a mechanical response in BMGs.

The basic idea of deformation defects characterization is to compare the structure of BMGs in the activated defects state with the structure expected for affine deformation i.e. ideal elastic state. It has been experimentally proved that localized inelastic deformation can be activated in elastic regime[14]. Thus, to examine the structure of the activated defects state *in-situ* diffraction studies were performed under uniaxial stress below yielding point. The two-dimensional (2-D) diffraction patterns were collected using high energy X-ray as a function of the applied load. By processing this 2-D diffraction pattern (see methods section), we obtain isotropic PDF, $g_0^{\,0}(r)$, reflecting volume change, and anisotropic PDF, $g_2^{\,0}(r)$, containing information about the shear-induced structural change. Since inelastic deformation in activated defects is purely a shear-induced event, $g_2^0(r)$ should provide intrinsic information about deformation defects. For pure affine-deformation state $g_{2\,\text{aff}}^0(r)$ can be calculated through an expression in terms of the first derivative of $g_0^0(r)$ as

$$g_{2\,\text{aff}}^0(r) = \pm\varepsilon_\text{aff}\frac{1}{\sqrt{5}}\frac{2(1+\nu)}{3}r\frac{dg_0^0(r)}{dr}$$

where "$\pm$" signs indicate compression and tension respectively, and $\varepsilon_\text{aff}$ is the affine strain. Therefore, structural comparison between the activated defects state and affine-deformation state is just the difference in the respective anisotropic PDFs.

Figure 1 compares experimental $g_2^0(r)$ for $Zr_{65}Cu_{17}Ni_8Al_{10}$ BMG under 400MPa at room temperature (300K) and corresponding calculated $g_{2\,\text{aff}}^0(r)$. Intriguingly, $g_2^0(r)$ deviates from $g_{2\,\text{aff}}^0(r)$ below 4 Å, which means $g_2^0(r)$ cannot be fitted by a constant elastic strain. The smaller magnitude of $g_2^0(r)$ below 4 Å indicates that local atomic strain is smaller than macroscopic elastic strain. The origin of the smaller local strain is examined by Iwashita et al.[16] using molecular dynamics (MD) simulation. Their results demonstrate that local topological rearrangement in terms of bond-breaking and bond-forming relaxes local stress leading to a reduced local strain. Consequently, the difference between $g_{2\,\text{aff}}^0(r)$ and $g_2^0(r)$, $\Delta g_2^0(r)$, reflects the intrinsic properties of deformation defects, and quantitative analysis of $\Delta g_2^0(r)$ is discussed in detail below.

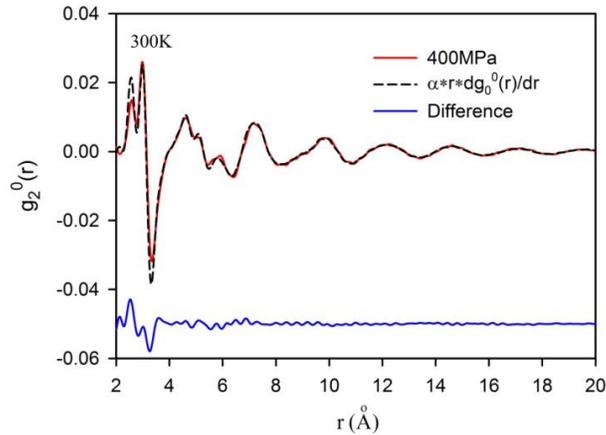

**Figure 1. Anisotropic PDFs for activated deformation and affine deformation state.** $g_2^0{}_{aff}(r)$ is calculated by matching $g_2^0(r)$ and $r * dg_0^0(r)/dr$ over $r$ range between 6.6 Å and 24 Å. Obvious difference between $g_2^0{}_{aff}(r)$ and $g_2^0(r)$ is seen below 4 Å.

First, an average radius of deformation defects can be determined from $\Delta g_2^0(r)$. Fig. 2 shows $\Delta g_2^0(r)$ for $Zr_{65}Cu_{17}Ni_8Al_{10}$ BMG under different stresses below yielding at RT. The amplitude beyond 4 Å is negligible, however, a dramatic difference below 4 Å demonstrates that topological rearrangements mainly occur in a spherical region with a radius about 4 Å, corresponding to the first atomic shell determined from $g_0^0(r)$ (see Supplementary Information, Figure S1). Accordingly, for $Zr_{65}Cu_{17}Ni_8Al_{10}$ BMG the deformation defects activated at RT includes 15 atoms, equal to the center atom plus an average coordination number (CN) calculated from $g_0^0(r)$ using

$$N_{CN} = \int_0^{r_{min}} 4\pi r^2 \rho_0 g_0^0(r) \, dr$$

where $r_{min}$ means the first minimum of $g_0^0(r)$ and $\rho_0$ is the atomic number density. More importantly, Fig. 2 reveals that there is no stress dependence for the defect size at RT.

Next, the amplitude of $\Delta g_2^0(r)$ reflects the other intrinsic characteristic of deformation defects, namely the density. Iwashita et al.[18] found that the number of atoms participating in topological rearrangement can be calculated by a formula given as

$$\Delta N_2^0 = N_2^0{}_{aff} - N_2^0$$
$$= \pm \int_0^{r_{min}} 4\pi r^2 \rho_0 \, g_2^0{}_{aff}(r) \, dr - \pm \int_0^{r_{min}} 4\pi r^2 \rho_0 \, g_2^0(r) \, dr$$
$$= \pm \int_0^{r_{min}} 4\pi r^2 \rho_0 \, \Delta g_2^0(r) \, dr$$

where "+" is for uniaxial tension case and "−" is for uniaxial compression. Using this equation, we found that $\Delta N_2^0$ linearly increase with stress in elastic regime (Fig. 2). Because defect size does not change with stress, the increase of $\Delta N_2^0$ must be due to the increase in number of activated defects. Clearly, our analysis based on $\Delta g_2^0(r)$ reveals that defect size and density can be directly characterized without a need to arbitrary tune several parameters in model methods.

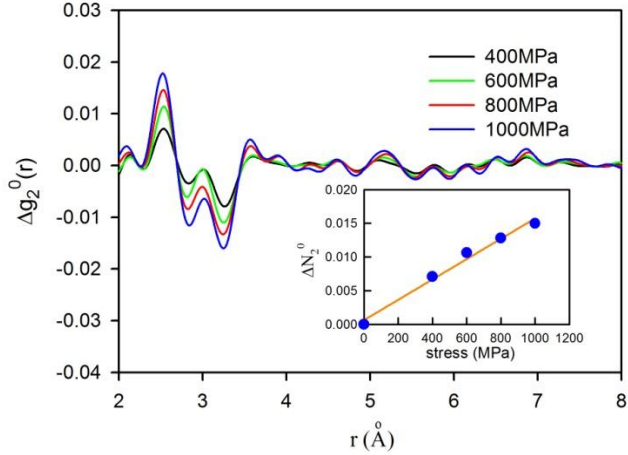

**Figure 2. Structural change induced by topological re-arrangement.** $\Delta g_2^0(r)$ is calculated for $Zr_{65}Cu_{17}Ni_8Al_{10}$ BMG under stresses without yielding. Defect size is determined from strong signal below 4 Å in $\Delta g_2^0(r)$. Defect density is reflected by $\Delta N_2^0$.

We now consider the effect of the chemical composition on defect size and density utilizing our method. Several studies suggest that the ductility of BMGs is correlated with high Poisson's ratio, but some abnormalities point out that this correlation may not be general: first, annealing below the glass transition temperature, $T_g$, makes ductile BMGs brittle but hardly change $v$ (less than ±2%)[19]; next, the ductility of Zr-based BMGs shows strong composition dependence, i.e. a small composition variation without a big change in $v$ leads to a brittle-ductile transition; third, Pd-based BMGs have higher $v$ than most Zr-based BMGs but they are brittle. To understand what controls intrinsic ductility we study several Zr-based BMGs with different ductility, Pd-, and Pt-based BMGs with the largest value $v$.

Fig. 3a and 3b show the composition dependence of $\Delta g_2^0(r)$ for Zr-, Pd-, and Pt-based BMGs respectively. All Zr-based BMGs show strong signal in the first atomic shell and small oscillations comparable with noise beyond the first atomic shell, same as stress-dependence behavior. Also, Pd- and Pt-based BMGs exhibit the same behavior although the second shell of Pt-based BMG has a relatively large oscillation. Comparing with Zr-based BMGs, Pd- and Pt-based BMGs have a smaller first atomic shell (around 3.5 Å), but the average CN of Pd and Pt BMGs is the same as Zr-based BMGs. Therefore, the deformation defects of Zr-, Pt-, and Pd-based BMGs include the same amount of atoms although the radius of deformation defects varies with composition. Our results, however, contradict the evaluation made by nanoindentation technique that BMGs with high $v$ have relatively large STZ size[7]. Moreover, annealing does not effectively change defect size for $Zr_{65}Cu_{17}Ni_8Al_{10}$ BMG, consistently with recent nanoindentation results by Li et al[8].

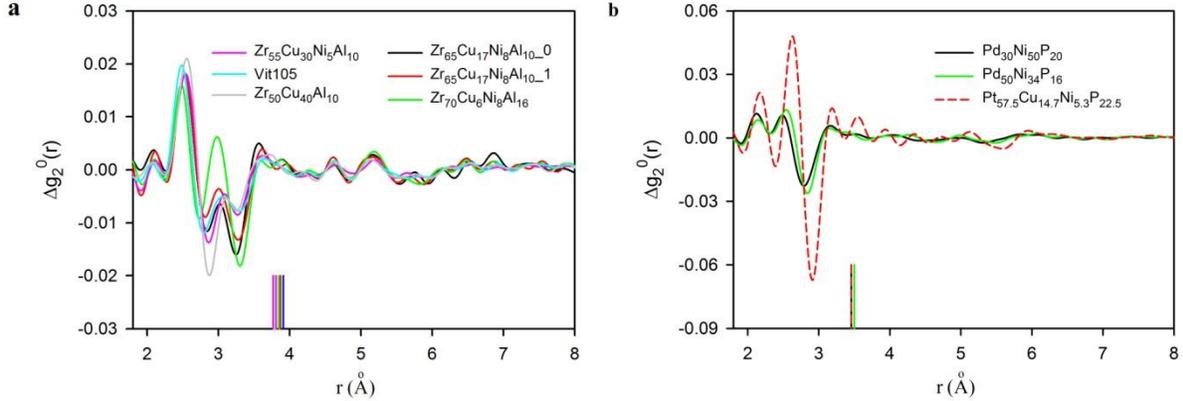

**Figure 3. Composition dependence of STZ size. a**, $\Delta g_2^0(r)$ for Zr-based BMGs. 0 and 1 refer to as-cast sample and sample annealed at 573K for 24 hrs. **b**, $\Delta g_2^0(r)$ for Pd-based BMGs and $Pt_{57.5}Cu_{14.7}Ni_{5.3}P_{22.5}$ BMG. Anisotropic PDF used to obtain $\Delta g_2^0(r)$ are all measured under compressive uniaxial stress, 1000MPa.

Now we examine the correlation between ductility and defect density. Here, instead of using stress dependent $\Delta N_2^0$, we defined a normalized parameter, $Y = \Delta N_2^0/N_2^0{}_{\text{aff}}$. The Y has a remarkable simple relationship with the concentration of atoms having topological change, $c_{TR}$, and the stress relaxed by topological rearrangement in activated defects, $\Delta\sigma$, given as

$$Y \propto \frac{c_{TR}}{\varepsilon_{\text{aff}}} \propto \frac{\Delta\sigma}{\sigma_{\text{aff}}}$$

where $\sigma_{\text{aff}}$ corresponds to affine deformation (see the derivation of this expression in Methods section). This expression clearly defines the physical meaning of Y, i.e. stress relaxation ability. The parameter Y does not depend on stress in elastic regime (see Supplementary Information, Figure S2).

Fig. 4 presents Y-$v$ map for all BMGs we studied. In this map, three BMGs above the orange line exhibit above 20% plasticity[12,20,21] and even show some tensile plasticity[12,22] while BMGs below the line have no tensile and compressive plasticity or compressive plasticity less than 2% (see stress-strain curves in Supplementary Information). This Y-$v$ map explains three anomalies mentioned above. For ductile $Zr_{65}Cu_{17}Ni_8Al_{10}$, annealing decreases its Y resulting in ductile-to-brittle transition. Two ductile Zr-based BMGs with the high concentration of Zr element ($\geq$ 65 at%) have their Y above the boundary. Brittle, Pd-based BMGs with $v$ higher than these two ductile Zr-based BMGs, however, have Y below the boundary. Interestingly, the Y of $Pt_{57.5}Cu_{14.7}Ni_{5.3}P_{22.5}$ is extraordinarily high, consistent with its resistance to induced brittleness by annealing[23]. Based on our characterization method, we reveal that defect density controls the ductility rather than defect size and $v$. Also it is clear from the map that Zr-based alloys are marginally ductile thus sensitive to production conditions and quenching rate.

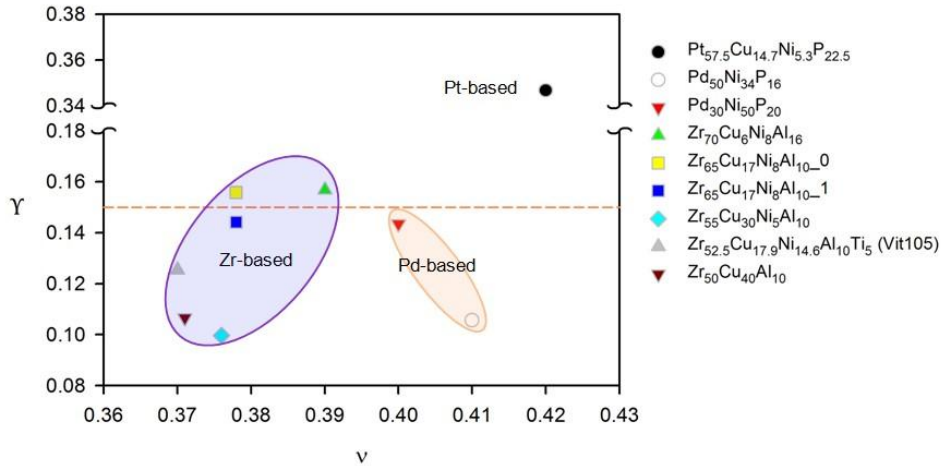

**Figure 4. Composition dependence of stress relaxation ability.** For each sample, Y is an average value of four measurements under different compressive stresses, 400MPa, 600MPa, 800MPa and 1000MPa. The orange dash line with Y = 0.15 is plot to emphasize three extremely ductile samples.

BMGs' plastic deformation at room temperature is asymmetrical under compressive and tensile load and the brittle fracture under tension is ascribed to pressure or normal stress effect. However, the influence of normal stress on the activation of deformation defects remains elusive. Here, we compare tension and compression cases using the above-mentioned method. By comparing $\Delta g_2^0(r)$, we find that the size of activated defects has no pressure dependence and topological rearrangement stays within the first atomic shell for both compression and tension (see

Supplementary Information, Figure S4). However, Y shows strong dependence on the sign of normal stress, as presented in Table 1. A small Y indicates BMGs' brittle tendency under tensile load. In addition, two intrinsic features of deformation defects in BMGs are discovered: a) deformation defects behave more unstable under tensile load; b) volume expansion does not assist local topological rearrangement in defects.

**Table 1. Normal stress effect on stress relaxation ability**

| Sample | $Zr_{50}Cu_{40}Al_{10}$ | $Zr_{55}Cu_{30}Ni_5Al_{10}$ | $Zr_{65}Cu_{17}Ni_8Al_{10}\_0$ | $Zr_{65}Cu_{17}Ni_8Al_{10}\_1$ |
|---|---|---|---|---|
| **Compression** (%) | 10.7 | 10.0 | 15.6 | 14.4 |
| **Tension** (%) | 7.1 | 7.9 | 10.3 | 8.0 |

We now discuss how temperature affects defects. Fig. 5 shows $g_2^0(r)$ measured at two temperatures (300K and 623K) for $Zr_{65}Cu_{17}Ni_8Al_{10}$ BMG. Importantly, $g_2^0(r)$ measured at 623K starts to deviate from $g_2^0{}_{aff}(r)$ at $r$ value beyond the first atomic shell, indicating defect size grows with the increase of temperature. Likewise to defect size, the value of Y also rises with temperature (Fig. 5e). The increase of Y with temperature is not linear but surges at temperature close to $T_g$. These experimental data demonstrate the atomic mechanism of inhomogeneous-to-homogeneous deformation transition proposed by the STZ theory. During homogeneous deformation STZs are no longer isolated but connect with each other[2], grow in size and stress is relaxed. Here, we only perform quasi-static mechanical testing during structural study, but it is worth to mention that defect size has strain-rate dependence at high temperature as found by MD simulation[16].

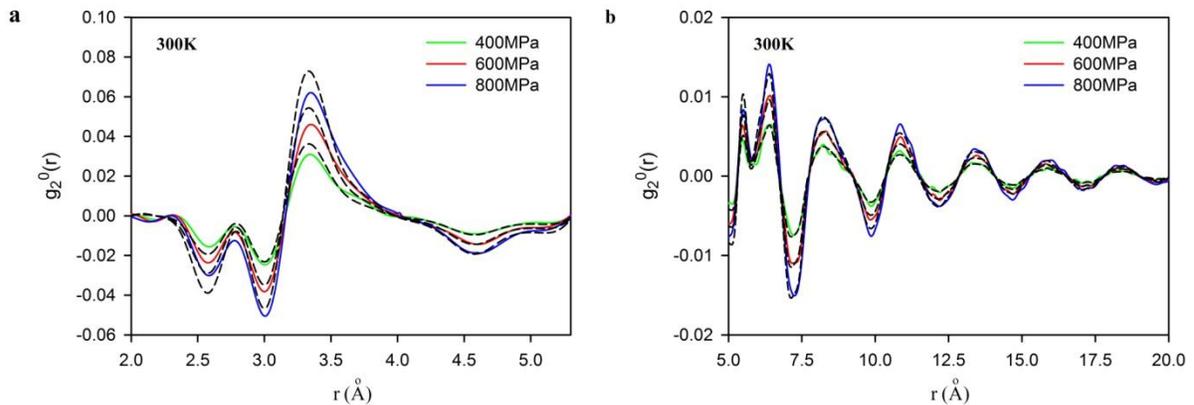

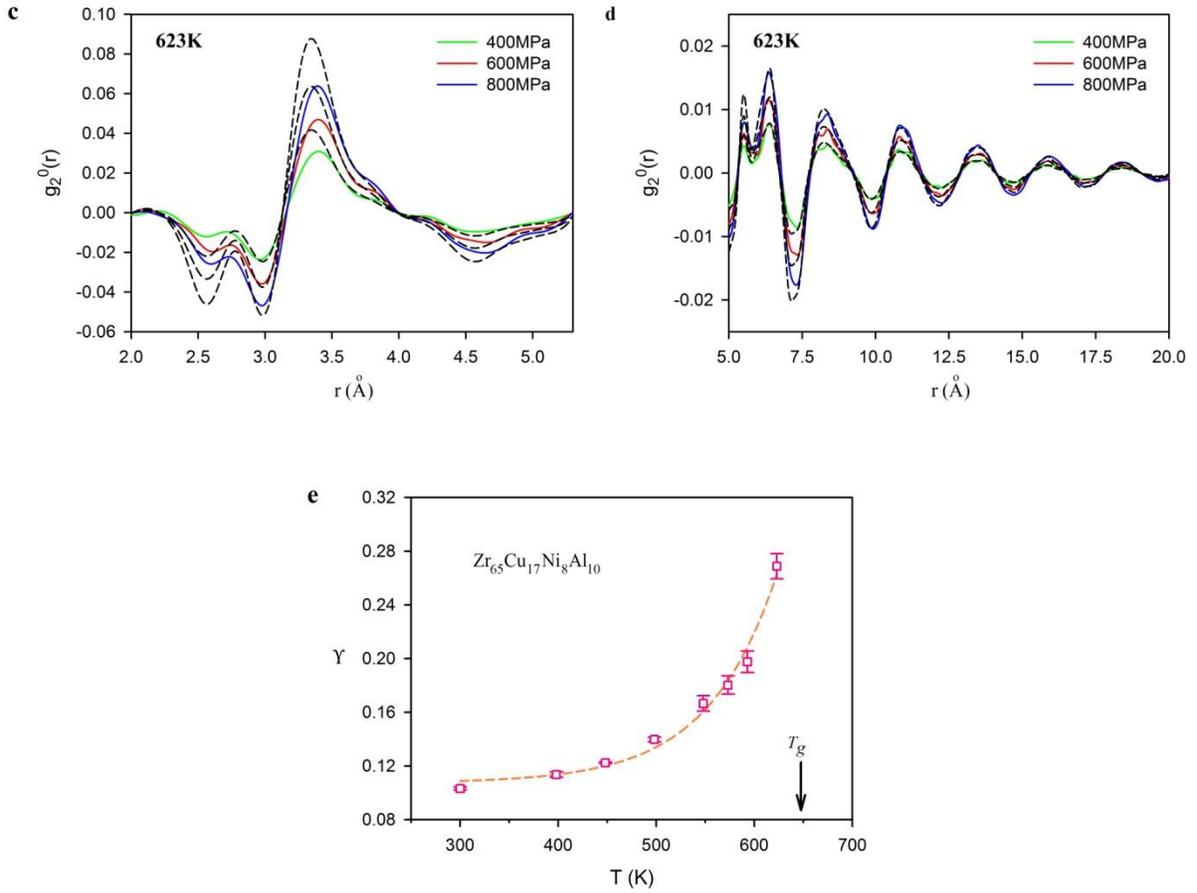

**Figure 5. Temperature effect on deformation defects. a** and **b**, Comparison of $g_2^0(r)$ and $g_2^0{}_{aff}(r)$ for $Zr_{65}Cu_{17}Ni_8Al_{10}$ BMG under different tensile stresses at 300K. **c** and **d**, Comparison of $g_2^0(r)$ and $g_2^0{}_{aff}(r)$ at 623K. In **a**, **b**, **c**, and **d**, colored curves are $g_2^0(r)$ and black dash lines are $g_2^0{}_{aff}(r)$. **e**, temperature effect on stress relaxation ability.

In the above study, we established a general characterization methodology to quantify deformation defects and found several important correlations with mechanical behavior of BMGs. Comparing with indirect characterization methods of STZ our method is parameter-free and can establish defect size and fractional density. Our results confirm the physical veracity of the phenomenological STZ theory at microscopic level, but the constitutive equations used in indirect STZ characterization methods need to be modified: at RT, the constitutive equation of plasticity should include the parameter, STZ density; at HT, two parameters, STZ size and density, are required.

**Methods**

**Sample preparation.** BMG rods were prepared by the tilt casting method[24]. The amorphous nature of the samples was confirmed by high energy X-ray and differential scanning calorimetry. For in-situ high energy X-ray diffraction study, cylinders with 2 mm diameter × 4 mm height were used for compression tests while standard dog-bone specimens 0.6 mm thick and 9 mm tall were used for tension tests.

**In-situ high energy X-ray diffraction.** Structural response to external load was studied by in-situ high energy X-ray diffraction at the 1-ID and 11-ID beam line of the Advanced Photon Source, Argonne National Laboratory. Infrared heater was used for HT study. The incident beam energy was 100 keV ($\lambda$ = 0.1239 Å). For Pt-based BMG, beam energy was tuned to 77keV ($\lambda$ = 0.161 Å) to reduce absorption. A stationary area detector having 2048 × 2048 pixels with 200 × 200 $\mu m^2$ pixel size was used to collect data. The detector was placed ~35 cm behind the sample. Calibration was performed using the $CeO_2$ powder standard. FIT2D software[25] was used to correct the beam polarization and the dark current. Under uniaxial tension or compression, BMGs' structure is no longer isotropic, and spherical harmonic expansion has to be performed to separate anisotropic part ($l=2$, $m=0$; other high order terms are negligible) and isotropic part ($l=0$, $m=0$) from the total PDF,

$$g(\boldsymbol{r}) = \sum_{l,m} g_l^m(r) Y_l^m(\frac{\boldsymbol{r}}{r})$$

where $Y_l^m$ is spherical harmonics. Details are in refs 13-15.

**Stress relaxation ability.** According to MD simulation, Suzuki et al. and Iwashita et al.[18] found a universal correlation between stress drop due to topological rearrangement and $c_{TR}$ and $\Delta N_2^0$:

$$\Delta\sigma = \sigma_{\text{aff}} - \sigma = \kappa E_\infty c_{TR} = \beta \Delta N_2^0$$

where $\kappa$ and $\beta$ are constants, $E_\infty$ is high frequency (instantaneous) Young's modulus. In order to study composition dependence, we normalize above equations using $\sigma_{\text{aff}}$ as

$$\frac{\Delta\sigma}{\sigma_{\text{aff}}} = \frac{\kappa E_\infty c_{TR}}{\sigma_{\text{aff}}} = \frac{\beta \Delta N_2^0}{\sigma_{\text{aff}}}$$

Using Hook's law, $\sigma_{\text{aff}} = \varepsilon_{\text{aff}} E_\infty$ and substituting into above equation, as

$$\frac{\kappa c_{TR}}{\varepsilon_{\text{aff}}} = \frac{\beta \Delta N_2^0}{\sigma_{\text{aff}}}$$

Additionally, since $\sigma_{\text{aff}}$ is proportional to $N_2^0{}_{\text{aff}}$, we obtain the following expression

$$\frac{\Delta N_2^0}{N_2^0{}_{\text{aff}}} \propto \frac{c_{TR}}{\varepsilon_{\text{aff}}}.$$

**Supplementary Information**

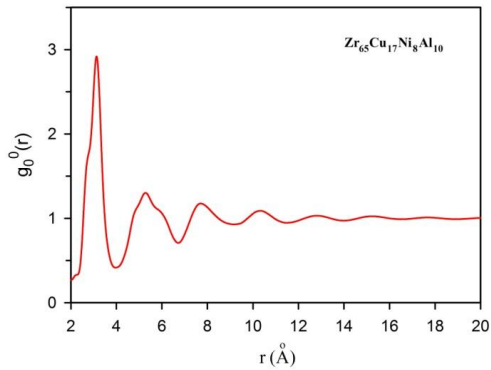

**Figure S1 Isotropic PDF for $Zr_{65}Cu_{17}Ni_8Al_{10}$.** The structure is measured at 0MPa and 300K.

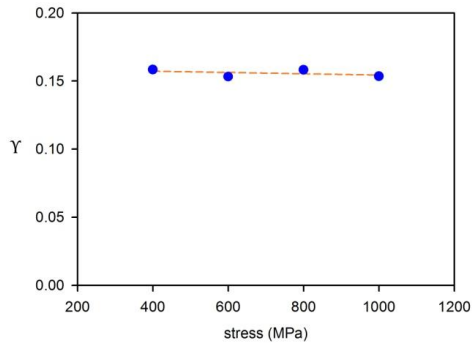

**Figure S2 The influence of stress on stress relaxation ability.**

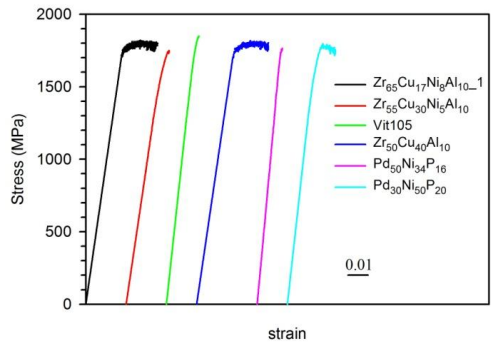

**Figure S3 Compressive stress-strain curves for Zr- and Pd-based BMGs.** Cylinder samples with 2mm diameter and 4mm length were used for uniaxial compressive tests at room temperature. The compression tests were performed at strain rate $2\times10^{-4}$ s$^{-1}$.

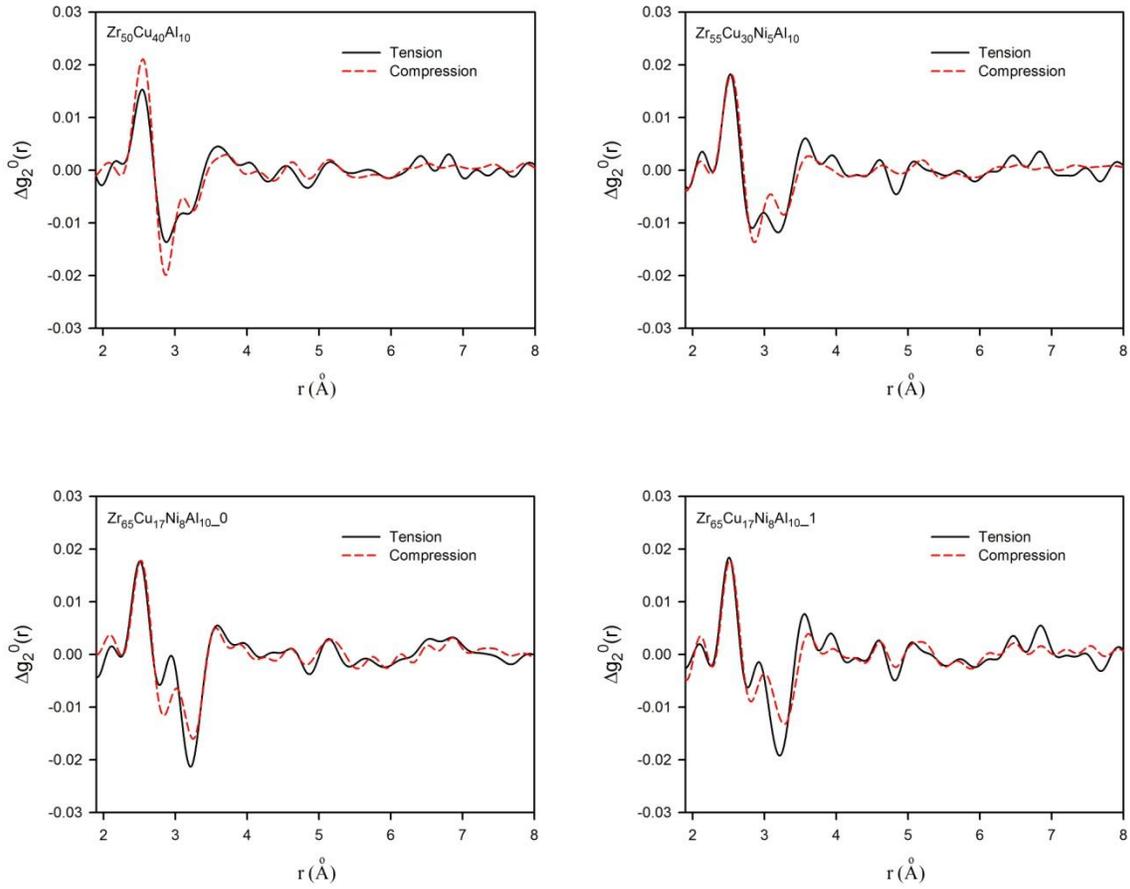

**Figure S4. Comparison of $\Delta g_2^0(r)$ under tensile and compressive load.** Anisotropic PDF used to obtain $\Delta g_2^0(r)$ are all measured under compressive or tensile uniaxial stress, 1000MPa. In order to do comparison, $\Delta g_2^0(r)$ for tension cases is flipped down.